# Ferromagnetic resonance and spin Hall magnetoresistance of $Tm_3Fe_5O_{12}$ films

Yufeng Wang, Peng Zhou[*], Shuai Liu, Yajun Qi[*], Tianjin Zhang[*]

Ministry of Education Key Laboratory for Green Preparation and Application of Functional Materials, Hubei Provincial Key Laboratory of Polymers, Collaborative Innovation Center for Advanced Organic Chemical Materials Co-constructed by the Province and Ministry, School of Materials Science and Engineering, Hubei University, Wuhan 430062, PR China

[*]E-mail addresses:

zhou@hubu.edu.cn (Peng Zhou),

yjqi@hubu.edu.cn (Yajun Qi)

zhangtj@hubu.edu.cn (Tianjin Zhang)

Ferromagnetic insulating garnet films with perpendicular magnetic anisotropy (PMA) are of great importance for the applications in spintronics. $Tm_3Fe_5O_{12}$ (TmIG) with magnetoelastic anisotropy has been proven to exhibit PMA more easily than $Y_3Fe_5O_{12}$. Here, magnetic parameters of TmIG films with various thicknesses are investigated by ferromagnetic resonance. The relationship between effective magnetic magnetization, surface perpendicular anisotropy field, magnetic anisotropy constants and film thickness is established. The parameters of spin Hall angle, spin diffusion length, and real part of the spin mixing conductance are obtained from the measurements of angular dependent magnetoresistance of TmIG/Pt heterostructures.

With the rapid development of spintronics, magnetic insulating nanometer thin films exhibiting perpendicular magnetic anisotropy (PMA) are becoming increasingly attractive. Compared with all-metallic devices, these insulating magnetic thin films with PMA enable non-volatile magnetic memory and logic devices with significantly reduced Joule heating[1]. Thulium iron garnet ($Tm_3Fe_5O_{12}$, TmIG) is a ferrimagnetic insulator that has been studied intensively due to its low Gilbert damping and robust PMA induced by magnetoelastic anisotropy [2-5].

Unlike the conventional method where magnetization is controlled by a magnetic field, the magnetization switching of (111)-oriented TmIG film was achieved due to the spontaneous breaking of mirror symmetry in magnetocrystalline anisotropy[6]. Similarly, the magnetization switching of TmIG/Pt heterostructure was realized without applying magnetic field because of the breaking of the in-plane magnetic symmetry[7]. In addition, the proximity-induced magnetic layer at the TmIG/Pt interface exhibited different impact on thermal-gradient-driven transverse voltage and electrically driven transverse resistance[8]. Magnons in magnetic insulator garnet films, such as $Y_3Fe_5O_{12}$ and TmIG, could be used to carry, transport, and process information[9-11]. Moreover, the interfacial Dzyaloshinskii-Moriya interaction in TmIG films capped with metals was responsible for the detection of topological spin textures[12, 13]. In fact, one of the most significant characteristics of TmIG is the presence of PMA, which has been confirmed in a film with a thickness of 5.6 nm[14]. The PMA has also been demonstrated to be associated with spin-current-induced magnetization switching[15], anomalous Hall hysteresis[16], film strain [17], nonlocal thermoelectric voltage[18], and Rashba spin-orbit coupling[19]. However, studying the fundamental magnetic parameters of TmIG films should serve as the foundation for the aforementioned researches.

In this study, we investigated the magnetic and electrical transport properties of (111)-oriented TmIG films with varying thicknesses using ferromagnetic resonance and angular dependent magnetoresistance. Very weak in-plane magnetic anisotropy was observed for the TmIG films. As the film thickness increased, the effective magnetization ($4\pi M_{eff}$) exhibited an upward trend, whereas the surface perpendicular anisotropy field ($H_p$) and magnetic anisotropy constants ($K_{p1}$ and $K_{p2}$) displayed a downward trend. In addition, spin Hall angle, spin diffusion length, and real part of the spin mixing conductance of 42 nm and 84 nm TmIG films were investigated.

TmIG films with thicknesses of 42 nm, 84 nm, 126 nm, and 168 nm, were grown on (111)-oriented $Gd_3Ga_5O_{12}$ (GGG) substrates by pulse laser deposition (PLD) with a KrF laser source (wavelength of 248 nm). The ablation energy and repetition rate were 300 mJ and 5 Hz, respectively. Substrate temperature and oxygen partial pressure were maintained 750 °C and 5 Pa, respectively. The Pt Hall-bar used for angular dependent magnetoresistance measurement was fabricated by sputtering, with length × width × thickness = 100 μm × 10 μm × 5 nm.

The structure of TmIG films was examined using X-ray diffraction (XRD, D8 Discover, Bruker, USA) with Cu Kα ($\lambda$ = 1.5406 Å) as the radiation source. The morphology of the cross-section and surface of the films was characterized using transmission electron microscopy (TEM, JEM-F200,

JOEL, USA) and atomic force microscopy (AFM, MEF-3D Origin, Asylum Research, USA), respectively. Magnetic hysteresis loops and angular dependent magnetoresistance were measured by physical property measurement system (PPMS, Quantum Design DynaCool, USA). Ferromagnetic resonance (FMR) was performed on an electron spin resonance spectrometry (EMX Plus, Bruker, USA). All measurements were conducted at room temperature.

The thickness of TmIG films is controlled by varying the duration of film growth. The cross-sectional TEM image of the thinnest film is shown in figure 1(b), where the thickness is approximately 42 nm. The thicknesses of the other films, specifically 84 nm, 126 nm, and 168 nm, were determined based on their respective growth times. The XRD spectrum of the 42 nm TmIG film, shown in figure 1(a), indicates that the film is grown along the [111] direction, with no other diffraction peaks detected. The calculated out-of-plane lattice spacing of $d^{TmIG,XRD}_{444}$ is 0.18 nm. Representative AFM image of TmIG film with thickness of 42 nm is shown in figure 1(c), revealing that the film surface is smooth, without showing obvious defects.

Figure 2 presents magnetic hysteresis loops of TmIG films, three points should be noted. (1) Magnetic anisotropy is observed to shift from PMA to in-plane anisotropy as the increase of film thickness. Specifically, PMA is exhibited by 42 nm and 84 nm TmIG films, while in-plane anisotropy is exhibited by the 168 nm film. The 126 nm film shows identical in-plane and out-of-plane magnetic hysteresis loops. (2) Saturation magnetization ($M_s$) for all of the TmIG films is approximately 180 emu/cm$^3$, which is higher than the early reported values[4, 19]. (3) Among the measured TmIG films, the 84 nm film shows the smallest in-plane coercive field of 65±5 Oe and the largest out-of-plane coercive field of 232±6 Oe.

The in-plane angular dependence of both FMR spectra and FMR field of a 42 nm TmIG film, as depicted in figure 3, does not exhibit significant angular variation, indicating a very weak in-plane anisotropy[1, 3]. Other films with different thicknesses display similar in-plane angular dependence in their FMR spectra (not shown here). To further investigate the magnetic parameters of TmIG films, we measured out-of-plane FMR spectra, as shown in figures 4(a) and S1. The out-of-plane angle $\theta_H$ denotes the orientation of bias magnetic field $H$ of FMR measurement with respect to the film normal, as illustrated in the inset of figure 4(c). Three important results are evident: (1) FMR field decreases as the increase of $\theta_H$ for TmIG films. (2) FMR field at $\theta_H = 0°$ is higher than the one at $\theta_H = 90°$ for 42 nm and 84 nm TmIG films, revealing that these two films do not exhibit PMA. This finding is inconsistent with the results obtained from magnetic hysteresis loops (as shown in figure 2). (3) For the 42 nm TmIG film, the difference in FMR fields between $\theta_H = 0°$ and $\theta_H = 90°$ is approximately 800 Oe, whereas for films with other thicknesses, this difference exceeds 2000 Oe.

Since the magnetization dynamics of TmIG films are described by the Landau-Lifshitz-Gilbert equation and the associated free energy density, the relationship between FMR field and $\theta_H$ can be fitted using the following equation [20-22].

$$\left(\frac{f}{\gamma}\right)^2 = [H\cos(\theta - \theta_H) - 4\pi M_{eff}\cos(2\theta) + H_p(3\sin^2\theta\cos^2\theta - \sin^4\theta)] \times [H\cos(\theta - \theta_H) - 4\pi M_{eff}\cos^2\theta + H_p\sin^2\theta\cos^2\theta] \quad (1)$$

Where $4\pi M_{eff} = 4\pi M_s - \frac{2K_{p1}}{M_s}$, $H_p = \frac{4K_{p2}}{M_s}$. $f$ is the FMR frequency (9.8 GHz in this work). $\theta$ is the angle between the equilibrium magnetization and the film normal.

Figures 4(b)-(e) show the fitted results using equation (1). Except for the 42 nm TmIG film, the experimental data for the 84 nm, 126 nm, and 168 nm films are fitted very well with the calculated FMR fields. The fitted parameters, including $4\pi M_{eff}$, $H_p$, $K_{p1}$ and $K_{p2}$, are exhibited in figure 5. Two points should be noted. (1) Gyromagnetic ratio $\gamma$ obtained from the fitting is 2.184 GHz/kOe, corresponding to a Landé g factor of 1.56. (2) With the increase of TmIG film thickness, $4\pi M_{eff}$ increases from 371 Oe to 1749 Oe, whereas the $H_p$ decreases from -571 Oe to -1008 Oe. (3) $K_{p1}$ and $K_{p2}$ for 42 nm TmIG film are $17.6\times10^4$ erg/cm$^3$ and $-2.6\times10^4$ erg/cm$^3$, respectively. Both values exhibit a downward trend as the film thickness increases.

Figure 6 displays the relationship between FMR linewidth and $\theta_H$ of TmIG films. The 42 nm TmIG film exhibits much broader linewidth than the thicker films. In general, FMR linewidth has three contributions, the intrinsic Gilbert linewidth broadening, the two-magnon scattering induced linewidth broadening, and the inhomogeneous linewidth broadening [20]. Specifically, the inhomogeneous linewidth broadening shows a larger contribution to the linewidth at $\theta_H = 0°$ than the contribution at $\theta_H = 90°$. Thus, the contribution from inhomogeneous linewidth broadening for 42 nm TmIG can be neglected, as shown in figure 6(a). In addition, the linewidth of the 168 nm film first increases and then decreases as $\theta_H$ increases. However, the linewidth of the other TmIG films is randomly distributed as $\theta_H$ increases.

Figures 7(d) and (e) show the angular dependent magnetoresistance of 84 nm and 42 nm TmIG films, respectively. Here, $\Delta R = R_{xx} - R_{min}$, $R_{xx}$ and $R_{min}$ are the resistance along $x$ (direction of current) and the minimum resistance during the angular variation, respectively. One can see that (1) $\Delta R$ for spin Hall magnetoresistance (SMR) is larger than the value for anisotropic magnetoresistance (AMR) and (2) $\Delta R$ for SMR of 84 nm is larger than the value of 42 nm film. The SMR ratio of TmIG/Pt can be described by the following equation[23].

$$\frac{R_{xx(90)} - R_{xx(0)}}{R_{xx(90)}} = \frac{\theta_{SH}^2 (2\lambda_{NM}^2 \rho_{NM}) t_{NM}^{-1} g_{\uparrow\downarrow} \tanh^2(\frac{t_{NM}}{2\lambda_{NM}})}{1 + 2\lambda_{NM}\rho_{NM} g_{\uparrow\downarrow} \coth(\frac{t_{NM}}{\lambda_{NM}})}, \quad (2)$$

Where NM in (2) represents normal metal (*i.e.*, Pt in this work). $R_{xx(0)}$ and $R_{xx(90)}$ are the resistance of TmIG/Pt at $\beta = 0°$ and $90°$, respectively. The parameters of $\theta_{SH}$, $\lambda_{NM}$, $\rho_{NM}$, $t_{NM}$, $g_{\uparrow\downarrow}$ are spin Hall angle, spin diffusion length, resistivity, thickness of Pt, the real part of the spin mixing conductance, respectively. $\lambda_{NM}$ and $\theta_{SH}$ are obtained via the following formulae[23, 24].

$$\lambda_{NM} = \frac{1}{\rho_{NM}} \times 0.63 \times 10^{-15} \tag{3}$$

$$\theta_{SH} = \frac{\sigma_{SH}}{(1/\rho_{NM})} \tag{4}$$

Where $\sigma_{SH}$ is spin Hall conductivity (we assume $\sigma_{SH} = 1.6 \times 10^5 \, \Omega^{-1} m^{-1}$ [24]). The measured $\rho_{NM}$ is 2.3×10⁻⁸ Ω·m. Thus, $\lambda_{NM}$ and $\theta_{SH}$ are calculated to be 26.5 nm and 3.68×10⁻³, respectively. Based on equation (2), the calculated $g_{\uparrow\downarrow}$ is 1.53× $10^{14} \, \Omega^{-1} m^{-2}$. The difference of $g_{\uparrow\downarrow}$ for 84 nm and 42 nm TmIG films can be neglected in this work.

Two notes should be made. (1) The magnetic hysteresis loops of 42 nm and 84 nm TmIG films exhibit PMA, whereas the corresponding angular dependent FMR results reveal in-plane anisotropy. That could be due to the competition between magnetocrystalline anisotropy and strain-induced magnetoelastic anisotropy. (2) Obvious deviation between the calculated and experimental FMR data is shown in figure 4(b). The underlying mechanisms deserve further investigation in the future.

In conclusion, magnetic and electrical transport properties of TmIG films with various thicknesses have been investigated. TmIG films exhibit weak in-plane anisotropy. 4π$M_{eff}$ exhibits upward trend while $H_p$ shows downward trend as the film thickness increases. The parameters of 42 nm and 84 nm TmIG films, including spin Hall angle, spin diffusion length, the real part of the spin mixing conductance, are obtained from the angular dependent magnetoresistance measurements. The research findings of this work will provide a reference for the application of TmIG in spintronics.

Acknowledgements:

This research was funded by the National Natural Science Foundation of China (No. 12104137, No. 12374083, and No. 12474083), the China Postdoctoral Science Foundation (No. 2020M672315), the Program of Introducing Talents of Discipline to Universities ("111 Project", D18025) China, and the Program of Hubei Key Laboratory of Ferro- & Piezoelectric Materials and Devices (No. K202013).

Figure captions:

Figure 1 (a) XRD $\theta$-$2\theta$ scan, (b) cross-sectional TEM image, (c) AFM image of TmIG film with thickness of 42 nm

Figure 2 Magnetic hysteresis loops of TmIG films with different thicknesses, (a) 42 nm, (b) 84 nm, (c) 126 nm, (d) 168 nm

Figure 3 (a) FMR spectra of TmIG film with thickness of 126 nm, where the magnetic field is rotated in the film plane with step of 15°, (b) FMR field as a function of in-plane magnetic field direction, data is obtained from (a)

Figure 4 (a) FMR spectra of TmIG film with thickness of 42 nm, where the magnetic field is rotated out of the film plane with step of 10°, (b) – (e) dependence of out-of-plane FMR field on $\theta_H$ of TmIG films with different thickness, the symbol and line represent experiment data and calculated values using equation (1), respectively. Inset in (c) shows the schematic diagram of the definition of $\theta_H$

Figure 5 Film thickness dependence of $4\pi M_{eff}$ and $H_p$ (a), and $K_{p1}$ and $K_{p2}$ (b)

Figure 6 FMR linewidth of TmIG film with thickness of (a) 42 nm, (b) 84 nm, (c) 126 nm, (d) 168 nm

Figure 7 Schematic diagrams of angular dependent magnetoresistance of TmIG film, (a) AMR, (b) SMR, (c) SMR +AMR. (d) and (e) angular dependent magnetoresistance with a magnetic field of 5 kOe for 84 nm and 42 nm TmIG films, respectively

Figure 1

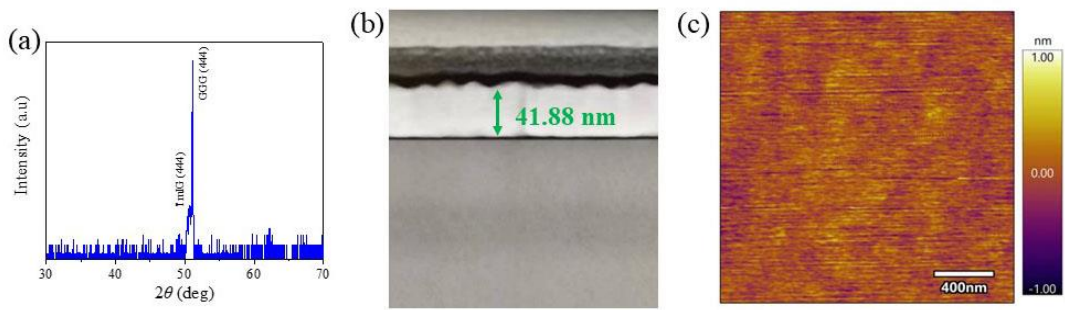

Figure 2

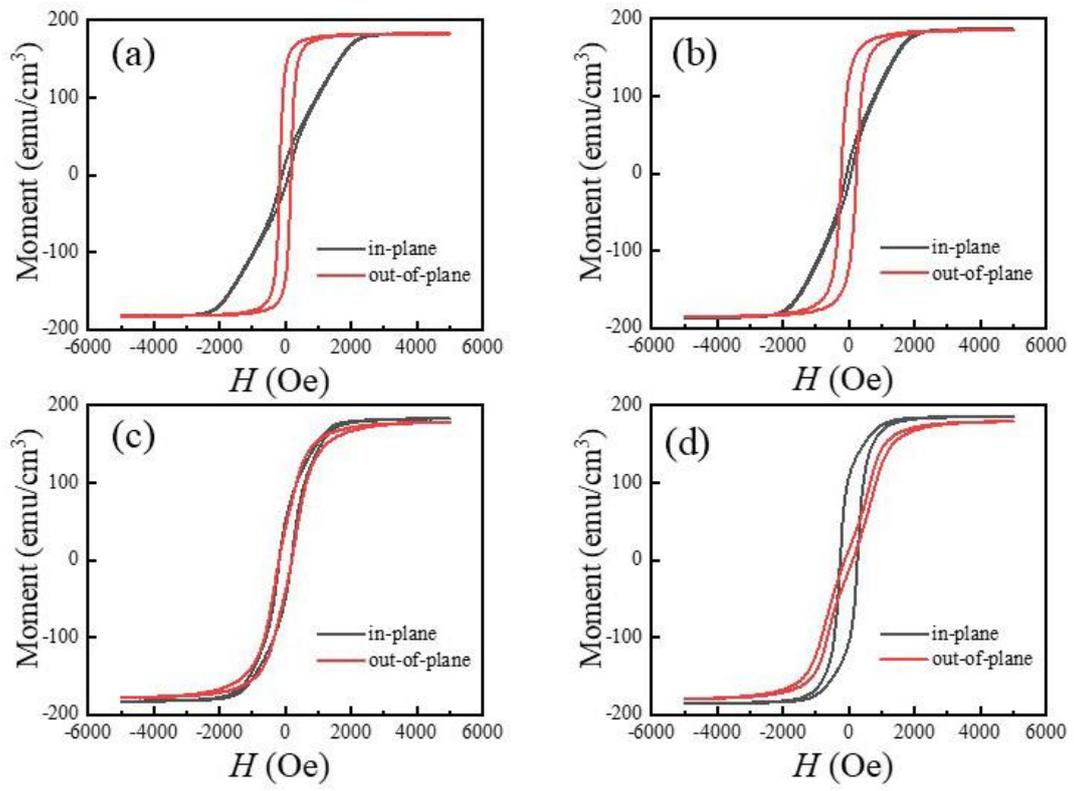

Figure 3

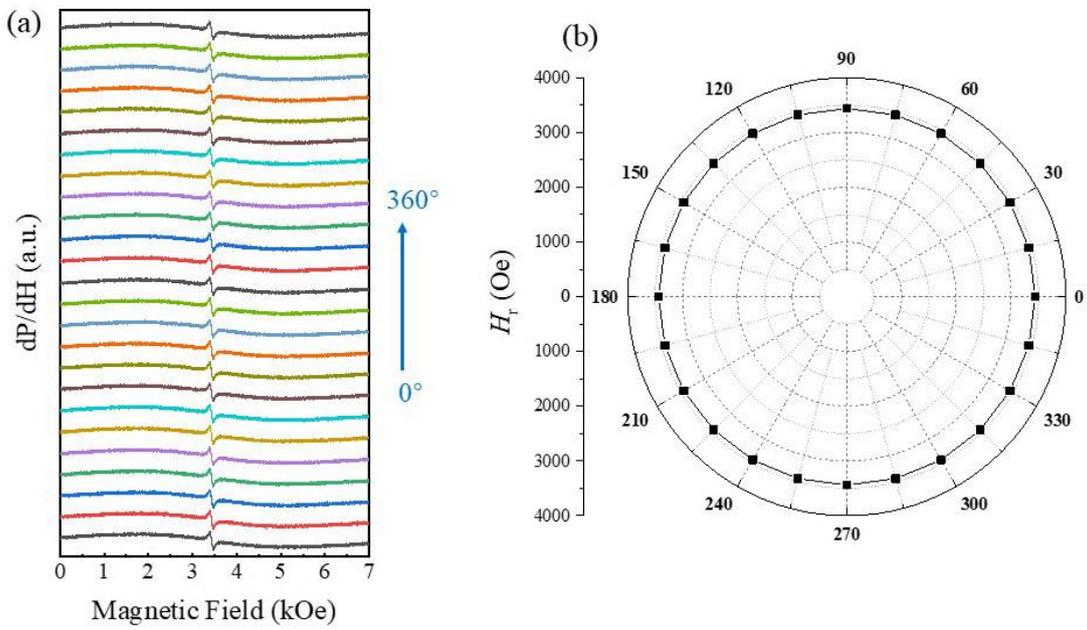

Figure 4

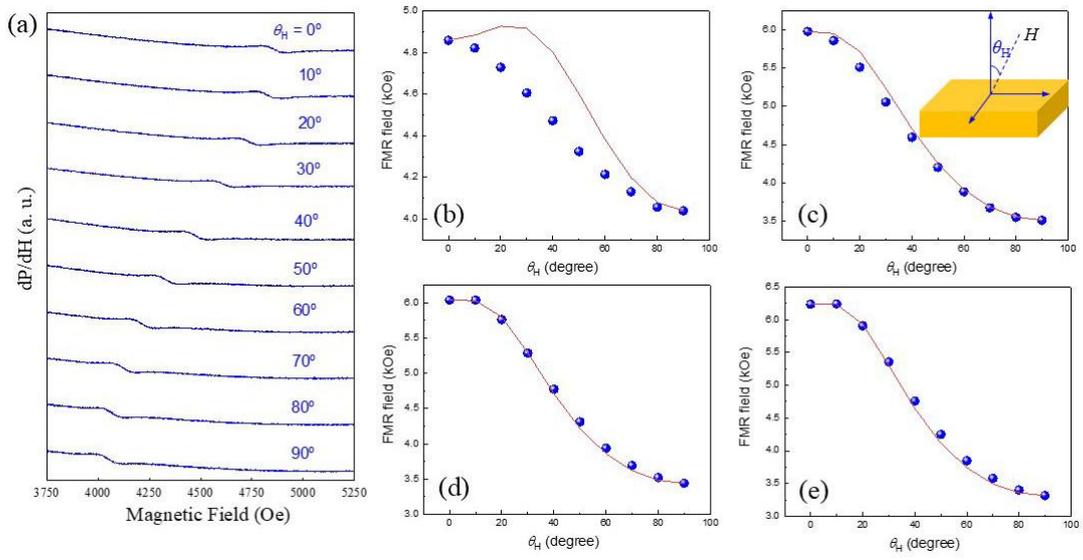

Figure 5

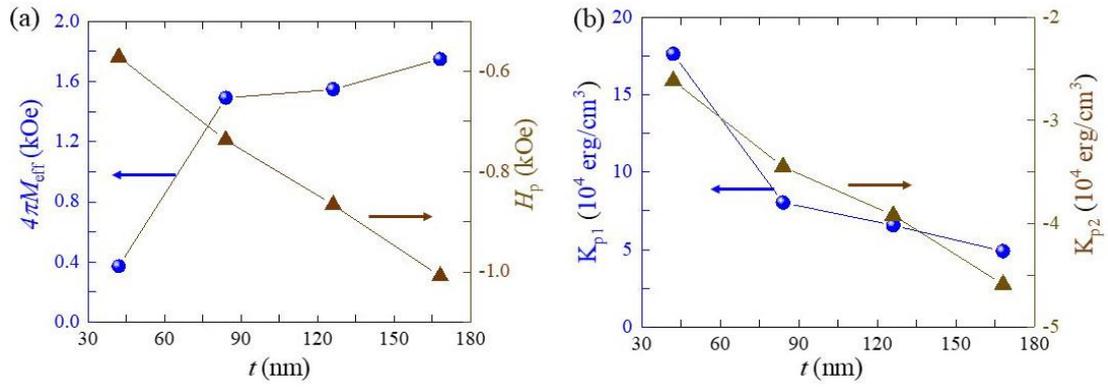

Figure 6

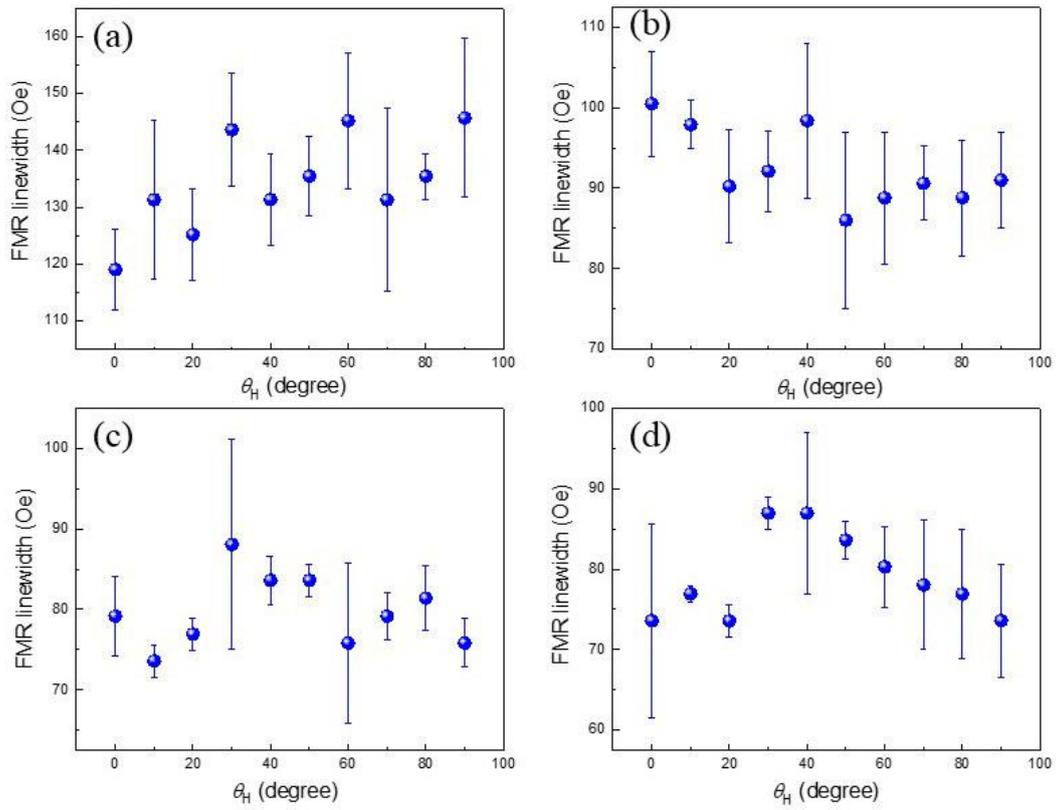

Figure 7

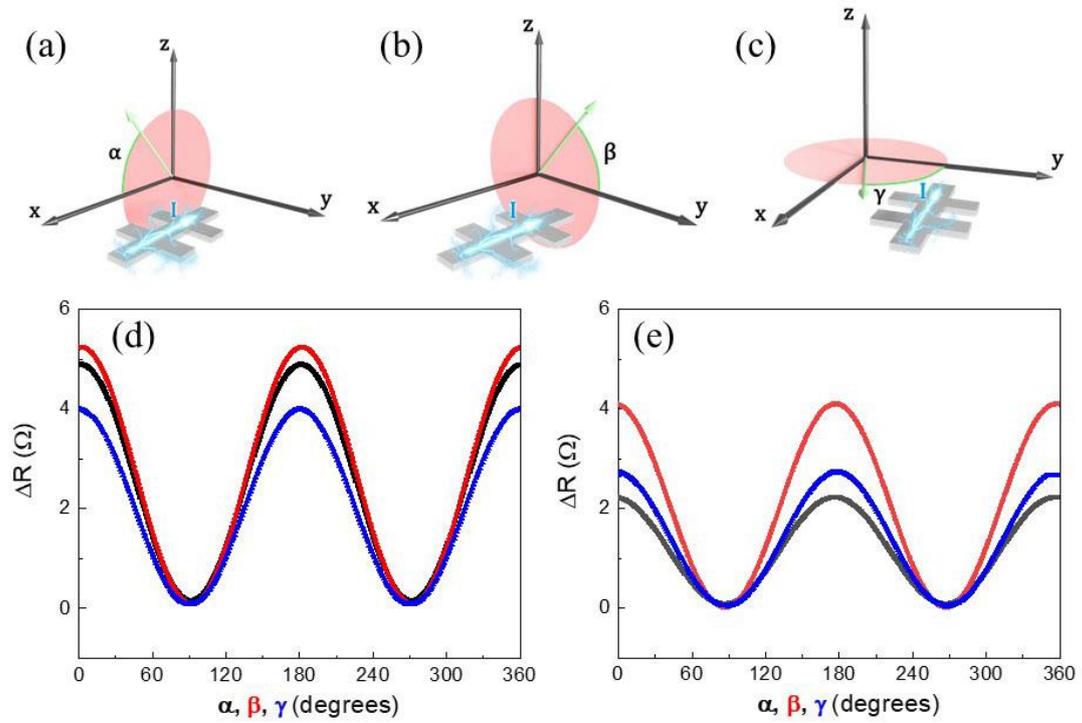

# Ferromagnetic resonance and spin Hall magnetoresistance of $Tm_3Fe_5O_{12}$ films

Yufeng Wang, Peng Zhou[*], Shuai Liu, Yajun Qi[*], Tianjin Zhang[*]

Ministry of Education Key Laboratory for Green Preparation and Application of Functional Materials, Hubei Provincial Key Laboratory of Polymers, Collaborative Innovation Center for Advanced Organic Chemical Materials Co-constructed by the Province and Ministry, School of Materials Science and Engineering, Hubei University, Wuhan 430062, PR China

[*]E-mail addresses:

zhou@hubu.edu.cn (Peng Zhou),

yjqi@hubu.edu.cn (Yajun Qi)

zhangtj@hubu.edu.cn (Tianjin Zhang)

Table SI Saturation magnetization ($M_s$), in-plane coercive field ($H_{c-in}$) and out-of-plane coercive field ($H_{c-out}$) of TmIG film with different film thickness. Values are obtained from figure 2

| Film thickness (nm) | $M_s$ (emu/cm$^3$) | $H_{c-in}$ (Oe) | $H_{c-out}$ (Oe) |
|---|---|---|---|
| 42 | 183±2 | 101±5 | 171±6 |
| 84 | 187±3 | 65±5 | 232±6 |
| 126 | 181±4 | 190±10 | 190±10 |
| 168 | 182±5 | 261±4 | 122±10 |

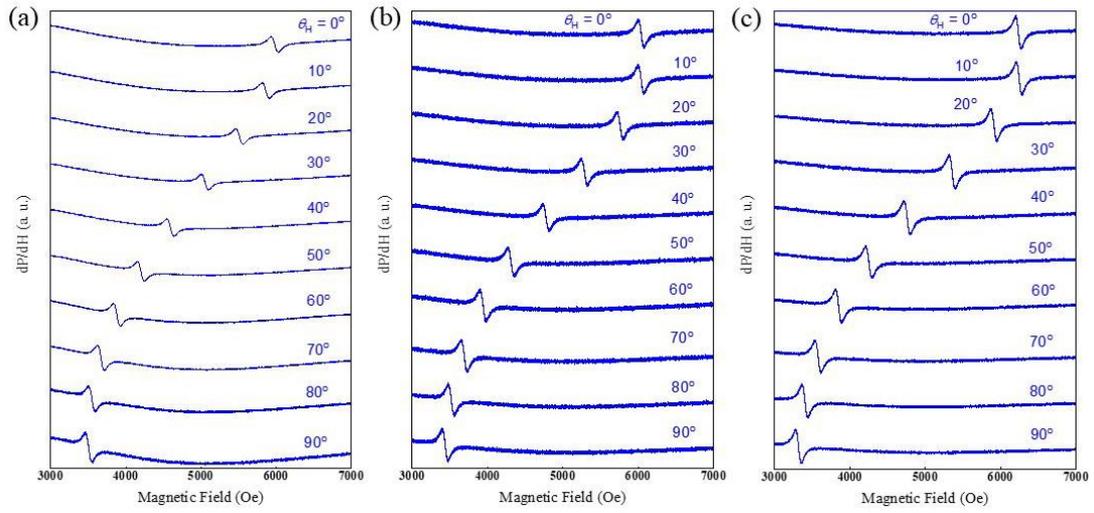

Figure S1 FMR spectra of TmIG films with thicknesses of 84 nm (a), 126 nm (b), and 168 nm (c), where the magnetic field is rotated out of the film plane with step of 10°